\documentclass[12pt]{article}
\usepackage{graphicx}
\usepackage{enumerate}
\usepackage{natbib}
\usepackage{url} % not crucial - just used below for the URL 

\usepackage{amsmath, amssymb}
\usepackage[colorlinks,citecolor=blue,urlcolor=blue]{hyperref}
\usepackage{enumitem}

\usepackage{multirow}
\usepackage{shortcuts}
\usepackage{graphicx}
\usepackage{nicefrac}
\usepackage{subcaption}
\usepackage{booktabs}
\usepackage{todonotes}
\usepackage[flushleft]{threeparttable}
\usepackage{bm}
\usepackage{bbm}
\usepackage[flushleft]{threeparttable}
\usepackage{float}
\usepackage{extarrows} 
\usepackage{adjustbox}

\usepackage{amsthm}

\usepackage{accents}
\newlength{\dhatheight}

\numberwithin{equation}{section}
\theoremstyle{plain}

\theoremstyle{definition}

\usepackage[ruled]{algorithm2e}

\SetArgSty{textrm}  % for algorithm2e
\usepackage{algorithmic}

\usepackage{multirow}

\newcommand{\blind}{1}

\begin{document}

\def\spacingset#1{\renewcommand{\baselinestretch}%
{#1}\small\normalsize} \spacingset{1}

%%%%%%%%%%%%%%%%%%%%%%%%%%%%%%%%%%%%%%%%%%%%%%%%%%%%%%%%%%%%%%%%%%%%%%%%%%%%%%

\if1\blind
{
  \title{\bf Kernel Optimal Orthogonality Weighting: A Balancing Approach to Estimating Effects of Continuous Treatments}
  \author{Nathan Kallus\\
    School of Operations Research and Information Engineering and \\
Cornell Tech, Cornell University, New York, New York 10044\\
\\
    Michele Santacatterina\thanks{
    Corresponding author. This material is based upon work supported by the National Science Foundation under Grants Nos. 1656996 and 1740822.}\hspace{.2cm}\\
    TRIPODS Center for Data Science for Improved Decision Making \\
    and Cornell Tech, Cornell University, New York, New York, 10044}
  \maketitle
} \fi

\if0\blind
{
  \bigskip
  \bigskip
  \bigskip
  \begin{center}
    {\LARGE\bf Kernel Optimal Orthogonality Weighting for Estimating Effects of Continuous Treatments}
\end{center}
  \medskip
} \fi

\bigskip
\begin{abstract}
Many scientific questions require estimating the effects of continuous treatments. Outcome modeling and weighted regression based on the generalized propensity score are the most commonly used methods to evaluate continuous effects. However, these techniques may be sensitive to model misspecification, extreme weights or both. In this paper, we propose Kernel Optimal Orthogonality Weighting (KOOW), a convex optimization-based method, for estimating the effects of continuous treatments. KOOW finds weights that minimize the worst-case penalized functional covariance between the continuous treatment and the confounders. By minimizing this quantity, KOOW successfully provides weights that orthogonalize confounders and the continuous treatment, thus providing optimal covariate balance, while controlling for extreme weights.  We valuate its comparative performance in a simulation study. Using data from the Women's Health Initiative observational study, we apply KOOW to evaluate the effect of red meat consumption on blood pressure.
\end{abstract}

\noindent%
\vfill

\newpage
\spacingset{1.45} % DON'T change the spacing!

%%%%%%%%%%%%%%%%%%%%%%%%%%%%%%%%%%%%%%%%%%%%%%%%%%%%%%%%%%%%%%%%%%%%%%%%%%%%%%%%%%%%%%%%%%%%%%%%%%%%%%%%%%%%%%%%%%%%%%%%%%%%%%%%%%%%%%%%%%%%%%%%%%%%%%%%%%%%%%%%%%%%%%%%%%%%%%%%%%%%%%%%%%%%%%%%%%%%%%%%%%%%%%%%%%%%%%%%%%%%%%%%%%%%%%%%%%%%%%%%%%%%%%%%%%%%%%%%%%%%%%%%%%%%%%%%%%%%%%%%%%%%%%%%%%%%%%%%%%%%%%%%%%%%%%%%%%%%%%%%%%%%%%%%%%%%%%%%%%%%%%%%%%%%%%%%%%%%%%%%%%%%%%%%%%%%%%%%%%%%%%%%%%%%%%%%%%%%%%%%%%%%%%%%%%%%%%%%%%%%%%%%%%%%%%%%%%
%%
%%
%%                  INTRODUCTION
%%
%%
%%%%%%%%%%%%%%%%%%%%%%%%%%%%%%%%%%%%%%%%%%%%%%%%%%%%%%%%%%%%%%%%%%%%%%%%%%%%%%%%%%%%%%%%%%%%%%%%%%%%%%%%%%%%%%%%%%%%%%%%%%%%%%%%%%%%%%%%%%%%%%%%%%%%%%%%%%%%%%%%%%%%%%%%%%%%%%%%%%%%%%%%%%%%%%%%%%%%%%%%%%%%%%%%%%%%%%%%%%%%%%%%%%%%%%%%%%%%%%%%%%%%%%%%%%%%%%%%%%%%%%%%%%%%%%%%%%%%%%%%%%%%%%%%%%%%%%%%%%%%%%%%%%%%%%%%%%%%%%%%%%%%%%%%%%%%%%%%%%%%%%%%%%%%%%%%%%%%%%%%%%%%%%%%%%%%%%%%%%%%%%%%%%%%%%%%%%%%%%%%%%%%%%%%%%%%%%%%%%%%%%%%%%%%%%%%%%

\section{Introduction}
\label{sec:intro}

%1. continuous important
%2. example
%3. obtain dr curve difficult because methods rely on correct
% gps or correct outcom model
%4. As a consequence, many researcher dichotomoized the trt 

The questions that motivate many scientific studies require estimating the effects of continuous treatments. Continuous treatments are usually indexed by doses and their relationships with the outcome are described by dose-response curves.  Consider, for instance, an observational study that aim at evaluating the relationship between red meat consumption and health outcomes such as high blood pressure, and cancer development. In this study, the amount of red meat that a person eats every day (dose in grams) is the continuous treatment under study, while the response is the person's blood pressure or probability of developing cancer. The medical literature on this topic is vast and it has shown that high red meat consumption may be harmful, leading to suggested levels of red meat consumption between 350 and 500 grams per week \citep{wrcf2018}. 

Common methods used to evaluate continuous effects are 1) outcome modeling, which first computes the regression function of the observed outcome on both the observed continuous treatment and the confounders and then takes its expectation over the
confounders \citep{flores2007estimation,florens2008identification,bia2011nonparametric}, and 2) methods based on the generalized propensity score (GPS) \citep{hirano2004propensity,imai2004causal} such as for example Inverse Probability Weighting (IPW) \citep{naimi2014constructing,robins2000marginal,cole2008constructing}. These methods rely on the correct specification of the outcome or GPS-model, respectively, which assumption is hardly ever met in any observational study. %Due to the complexity of the covariate-treatment relationships and the difficulty to evaluate covariate balance with continuous treatments, this issue exacerbates when estimating dose-response curves. 
In addition, when using IPW-based methods, weights can be extreme leading to erroneous inferences \citep{kang2007demystifying}.  %Consequently, although the natural object of interest of many studies is the dose-response curve, applied researchers categorize the continuous treatment in two or few categories. This leads to a loss of information and other related issues (see for instance \cite{cox1957note,cohen1983cost,fedorov2009consequences} on the issue of categorization). 
Many other methods have been developed and we review them in Section \ref{sec:rela_work}.

In this paper, rather than using outcome modeling or the GPS,  we propose Kernel Optimal Orthogonality Weighting (KOOW), a novel method that optimally finds weights that minimize the worst-case penalized functional covariance between the continuous treatment and the confounders. KOOW aims at eliminating any relationship between confounders and the continuous treatment, thus providing optimal covariate balance. To do so, in Section \ref{sec:KOOW} we start by proposing a functional formulation of the covariance between the continuous treatment and the confounders. We then define the penalized worst-case functional covariance and use kernels and quadratic programming to find the weights that minimize this quantity. Finally, for estimating the dose-response curve, we propose to plug-in the obtained optimal weights into a nonlinear weighted ordinary least squares estimator or a weighted local polynomial regression estimator.  We describe the properties of the proposed methodology and provide practical guidelines on its use in Section \ref{sec:prop} and Section \ref{sec:guidelines}, respectively. In Section \ref{sec:simu}, we report the results of a simulation study aimed at valuating the performance of KOOW in a variety of different scenarios. 
% by using kernels to model the functional relationship between the continuous treatment and the confounders, the proposed methodology has the quality of being highly flexible which consequently mitigates possible model misspecifications. 
%In addition, KOOW simply extend the concept of covariate balancing to continuous treatment targeting orthogonality while simultaneously penalizing extreme weights and thus controlling for precision. 
We apply the proposed methodology on the evaluation of red meat consumption on blood pressure among women of the Women Health Initiative observational study (Section \ref{sec:case_study}). We conclude with some remarks in Section \ref{sec:conc}.

\subsection{Related work}
\label{sec:rela_work}

In practice, one of the most common method used to evaluate continuous effects is outcome modeling, which directly models the relationship between the outcome, the treatment and confounders \citep{flores2007estimation}, \citep[Section 6.2]{hill2011bayesian}, \citep[Section 3.1]{zhang2016causal}. %However, it is well known that by relying entirely on the correct specification of the outcome model, these methods suffers from model misspecification.  Furthermore, these methods do not not incorporate any available information about the treatment mechanism.
Methods based on the GPS have been proposed.  
\cite{hirano2004propensity,imai2004causal} suggested to use an outcome model conditioned on the estimated GPS. \cite{lu2011optimal,lu2001matching} proposed matching on the GPS using a non-bipartite matching algorithm. More recently, \cite{wu2018matching} developed an new approach for GPS caliper matching. IPW-based methods, such as Marginal Structural Models (MSM)  \citep{robins2000marginal,cole2008constructing} have been extended in the case of continuous treatments \citep{naimi2014constructing,gill2001causal,zhang2016causal}. A variety of methods have been proposed to estimate the GPS.  \cite{zhu2015boosting} proposed a boosting algorithm for estimating GPS. \cite{kreif2015evaluation} suggested to use Super Learner \citep{van2003unified} for both the GPS and the outcome models. More generally, \cite{kennedy2017non} developed a nonparametric doubly robust estimator for estimating dose-response curves. When the estimated GPS is very close to zero or one, which in the case of a continuous treatment is inevitable, the corresponding obtained weights can be very extreme leading to erroneous inferences \citep{kang2007demystifying}. Methods have been proposed to overcome the issue of extreme weights \citep[among others]{santacatterina2017optimal,cole2008constructing,xiao2013comparison}. Methods that target covariate balance have been recently developed. \cite{fong2018covariate} extended the covariate balancing propensity score methodology to continuous treatment by developing both a parametric and a nonparametric version of the method. \cite{yiu2018covariate} introduced a general framework based on a constrained optimization problem that find weights that eliminate the relationship between the continuous treatment and the covariates. \cite{2019arXiv190101230A} introduced permutation weighting, which finds weights based on density-ratio estimation via probabilistic classification. %Partial mean methods have been developed. Partial mean follows a regression approach by first computing the regression function of the observed outcome onboth the observed continuous treatment and the confounders and then by taking its expectation over theconfounders \citep{flores2007estimation,florens2008identification,bia2011nonparametric}. 
The computer science and optimization literature has been also proposing novel methods. For example, see \cite{kallus2018policy,sondhi2019balanced,krishnamurthy2019contextual} for policy evaluation, and  \cite{demirer2019semi} for policy learning with continuous actions.  Other methods include that of \cite{galvao2015uniformly} in which  a semiparametric two-step estimator for estimating the dose-response function is proposed, and the extension of the G-computation \citep{robins1986new} methodology to continuous treatments \citep{neugebauer2006g,gill2004causal}. Methods have been proposed to estimate effects of continuous treatments in a variety of different setting, such as quantile continuous treatment effects \citep{alejo2018quantile}, difference-in-differences in repeated cross sections with continuous treatments \citep{d2013nonlinear}, optimal dynamic continuous treatment regimes \citep{barrett2014doubly}, and dose-response function for longitudinal data \citep{moodie2012estimation}. Finally, our methods builds upon some results on kernel methods for testing the independence of two random variables \citep{gretton2005kernel,gretton2007kernel}, and kernel mean matching for density-ratio estimation \citep{huang2007correcting,gretton2009covariate} \citep[Section 3.3]{sugiyama2012density}. 

Our main contribution relative to this body of work is to provide a general and intuitive balancing approach based on conventional optimization techniques to estimating effects of continuous treatments.

%%%%%%%%%%%%%%%%%%%%%%%%%%%%%%%%%%%%%%%%%%%%%%%%%%%%%%%%%%%%%%%%%%%%%%%%%%%%%%%%%%%%%%%%%%%%%%%%%%%%%%%%%%%%%%%%%%%%%%%%%%%%%%%%%%%%%%%%%%%%%%%%%%%%%%%%%%%%%%%%%%%%%%%%%%%%%%%%%%%%%%%%%%%%%%%%%%%%%%%%%%%%%%%%%%%%%%%%%%%%%%%%%%%%%%%%%%%%%%%%%%%%%%%%%%%%%%%%%%%%%%%%%%%%%%%%%%%%%%%%%%%%%%%%%%%%%%%%%%%%%%%%%%%%%%%%%%%%%%%%%%%%%%%%%%%%%%%%%%%%%%%%%%%%%%%%%%%%%%%%%%%%%%%%%%%%%%%%%%%%%%%%%%%%%%%%%%%%%%%%%%%%%%%%%%%%%%%%%%%%%%%%%%%%%%%%%%
%%
%%
%%                  KOOW
%%
%%
%%%%%%%%%%%%%%%%%%%%%%%%%%%%%%%%%%%%%%%%%%%%%%%%%%%%%%%%%%%%%%%%%%%%%%%%%%%%%%%%%%%%%%%%%%%%%%%%%%%%%%%%%%%%%%%%%%%%%%%%%%%%%%%%%%%%%%%%%%%%%%%%%%%%%%%%%%%%%%%%%%%%%%%%%%%%%%%%%%%%%%%%%%%%%%%%%%%%%%%%%%%%%%%%%%%%%%%%%%%%%%%%%%%%%%%%%%%%%%%%%%%%%%%%%%%%%%%%%%%%%%%%%%%%%%%%%%%%%%%%%%%%%%%%%%%%%%%%%%%%%%%%%%%%%%%%%%%%%%%%%%%%%%%%%%%%%%%%%%%%%%%%%%%%%%%%%%%%%%%%%%%%%%%%%%%%%%%%%%%%%%%%%%%%%%%%%%%%%%%%%%%%%%%%%%%%%%%%%%%%%%%%%%%%%%%%%%

\section{Kernel Optimal Orthogonality Weighting}
\label{sec:KOOW}

In this Section we present KOOW. We start by defining the functional covariance between the continuous treatment and the confounders (Section \ref{sec:cov_func}). Since this quantity depends on unknown functions, in Section \ref{sec:wc_cov_func}, we define the worst-case functional covariance. Weights that minimize this quantity may be extreme. In Section \ref{sec:wc_cov_func} we also add a penalization term to control for extreme weights. In Section \ref{sec:qp_wc}, by using kernels and standard results from Reproducing Kernel Hilbert Spaces (RKHS), we find the set of weights that minimize the worst-case penalized functional covariance by defining a linearly-constrained convex optimization problem.

%%%%%%%%%%%%%%%%%%%%%%%%%%%%%%%%%%%%%%%%%%%%%%%%%%%%%%%%%%%%%%%%%%%%%%%%%%%%%%%%%%%%%%%%%%%%%%%%%%%%%%%%%%%%%%%%%%%%%%%%%%%%%%%%%%%%%%%%%%%%%%%%%%%%%%%%%%%%%%%%%%%%%%%%%%%%%%%%%%%%%%%%%%%%%%%%%%%%%%%%%%%%%%%%%%%%%%%%%%%
%%
%%
%%                  Functional covariance
%%
%%
%%%%%%%%%%%%%%%%%%%%%%%%%%%%%%%%%%%%%%%%%%%%%%%%%%%%%%%%%%%%%%%%%%%%%%%%%%%%%%%%%%%%%%%%%%%%%%%%%%%%%%%%%%%%%%%%%%%%%%%%%%%%%%%%%%%%%%%%%%%%%%%%%%%%%%%%%%%%%%%%%%%%%%%%%%%%%%%%%%%%%%%%%%%%%%%%%%%%%%%%%%%%%%%%%%%%%%%%%%%%%%%%%%%%%%%%%%%%%%%%%%%%%%%%%%%%%%%%%%%%%%%%%%%%%%%%%%%%%%%%%%%%%%%%%%%%%%%%%%%%%%%%%%%%%%%%%%%%%%%%%

\subsection{Functional covariance}
\label{sec:cov_func}

\begin{figure}[H] 
\begin{center}
%\begin{adjustbox}{angle=0,width=\columnwidth,center}
\includegraphics[scale=0.7]{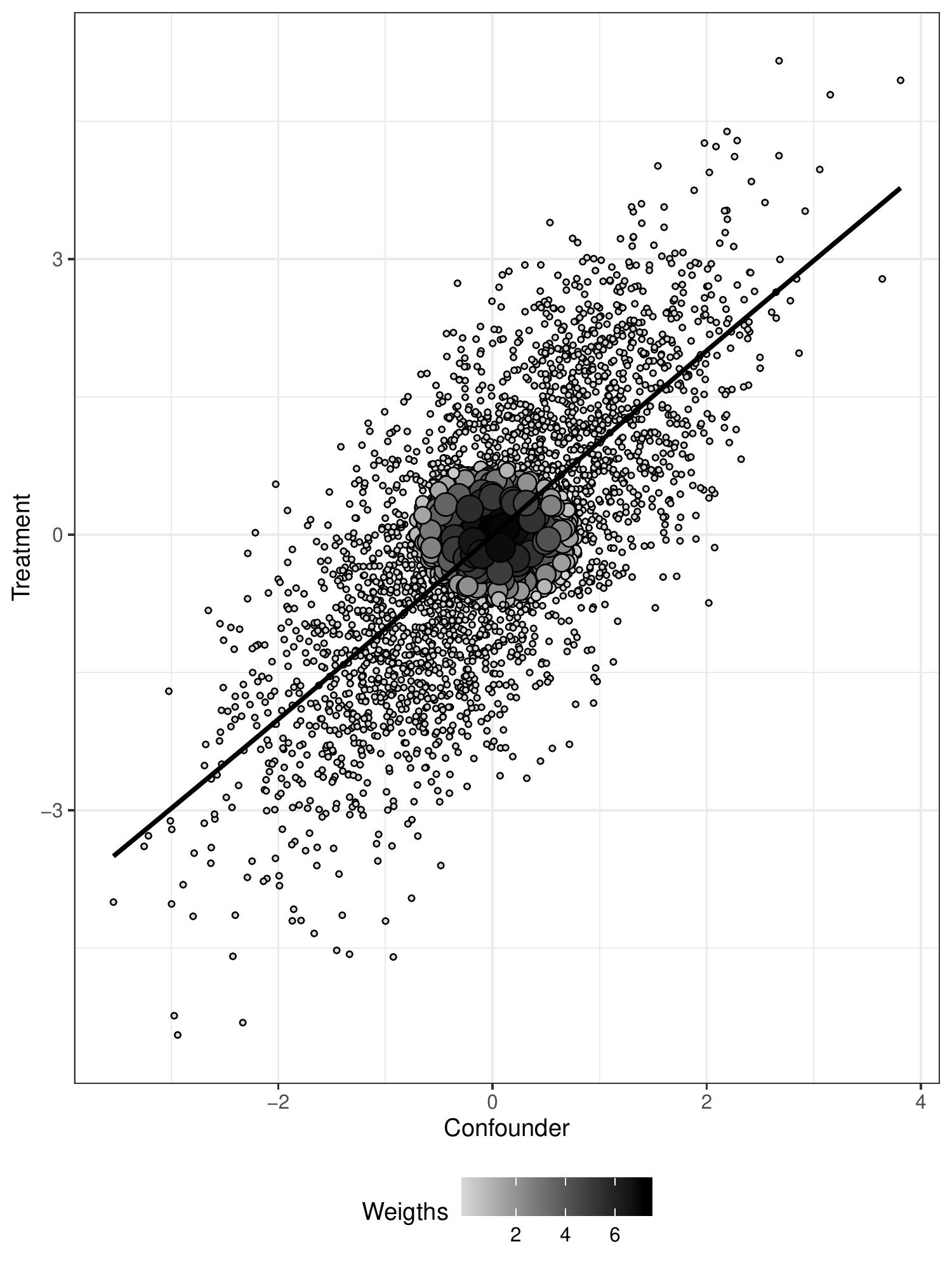}
%\end{adjustbox}
\end{center}
\caption{\footnotesize Scatterplot between a confounder and a continuous treatment. The size and the darkness of the circles refer to the size of the weights that minimize eq \eqref{eq:WFC}.
\label{fig:plot_expw} }

\end{figure}

Suppose we have a simple random sample with replacement of size $n$ from a population. For each unit $i$ in $1,\dots,n$ let $X_i$ and $A_i$ be the observed confounder and treatment value. Let $\Xs$ and $\As$ denote all the observed confounders and treatment values. The main idea is to find a set of weights that minimizes the following empirical functional covariance between confounders $\Xs$ and treatment $\As$. We define this quantity as, 

\begin{align}
\label{eq:WFC}
\delta(\Ws,f) = \nf \sum_{i=1}^n W_i f(X_i,A_i)-\nnf \sum_{i=1}^n \sum_{j=1}^n f(X_i,A_j),
\end{align}

\noindent
where $f(x,a)$ is an unknown function that describe the relationship between the continuous treatment and the confounders (here, $a$ and $x$ are just two dummy variables). Equation \eqref{eq:WFC} suggests finding weights that re-balance the joint distribution between treatment and confounders to be the same as that of the product of the two distributions. We now provide a straightforward example to clarify ideas. Figure \ref{fig:plot_expw} depicts the relationship between a normally distributed confounder $X$ with mean 0 and variance 1, and a normally distributed treatment $A$ with mean $X$ and unit variance. The straight line shows the relationship between the two while the dots represent the values obtained for the two variables. The darkness and the size of the circles represent the size of the weights, \ie, the darker/larger the circle, the larger the weight. A set of weights that minimize eq. \eqref{eq:WFC}, weights more those units with no relationship between the confounder and the continuous treatment making them orthogonal. In other words, the obtained weights eliminate any associations between the continuous treatment and the confounders, thus providing optimal covariate balance. As a matter of fact, recent literature on GPS-based methods \citep{austin2019assessing,zhu2015boosting} suggests the use of correlation-based diagnostics, such as the weighted correlation, to assess covariate balance. Our idea is not only to use this quantity as diagnostic but actually to find the set of weights that minimize it. 

%make the confounder and the continuous treatment orthogonal, \ie, uncorrelated. In other words, it
. 
%In other words, it provides a set of weights . In the next Section we provide more details on how to find these weights.

%This can be easily seen in Figure \ref{fig:plot_expw} by the darkness and size of the dots picked by the optimization problem. The  which indicates the usual uncorrelatedeness. We now provide details on how to obtain these weights.

%More details on how to find these weights are given in the next Sections. 

%%%%%%%%%%%%%%%%%%%%%%%%%%%%%%%%%%%%%%%%%%%%%%%%%%%%%%%%%%%%%%%%%%%%%%%%%%%%%%%%%%%%%%%%%%%%%%%%%%%%%%%%%%%%%%%%%%%%%%%%%%%%%%%%%%%%%%%%%%%%%%%%%%%%%%%%%%%%%%%%%%%%%%%%%%%%%%%%%%%%%%%%%%%%%%%%%%%%%%%%%%%%%%%%%%%%%%%%%%%
%%
%%
%%                  Worst-case penalized functional covariance
%%
%%
%%%%%%%%%%%%%%%%%%%%%%%%%%%%%%%%%%%%%%%%%%%%%%%%%%%%%%%%%%%%%%%%%%%%%%%%%%%%%%%%%%%%%%%%%%%%%%%%%%%%%%%%%%%%%%%%%%%%%%%%%%%%%%%%%%%%%%%%%%%%%%%%%%%%%%%%%%%%%%%%%%%%%%%%%%%%%%%%%%%%%%%%%%%%%%%%%%%%%%%%%%%%%%%%%%%%%%%%%%%%%%%%%%%%%%%%%%%%%%%%%%%%%%%%%%%%%%%%%%%%%%%%%%%%%%%%%%%%%%%%%%%%%%%%%%%%%%%%%%%%%%%%%%%%%%%%%%%%%%%%%

\subsection{Worst-case penalized functional covariance}
\label{sec:wc_cov_func}

% ----> Define the covariance $\delta(\Ws,f)$ as a linear operator. We define its Operator norm. 
The functional covariance showed in equation \ref{eq:WFC}, depends on the unknown function $f$. In this Section, we propose to minimize the worst-case functional covariance normalized by the magnitude of the function $f$. To do so, we start by embedding the function $f$ into a seminormed space  with seminorm (a norm that can also assign the values $0$ and $\infty$ to nonzero elements), $\| \cdot \|$. We first define the normalized worst-case functional covariance as
\begin{align}
\label{eq:wc_wfc}
\Delta(\Ws) &= \sup_{\substack{ f }} \frac{\delta(\Ws,f)}{\| f \|}  = \sup_{\substack{ \|f\| \leq 1 }} \delta(\Ws,f),
\end{align}

\noindent
which can be seen as the dual norm of the continuous linear operator, $\delta(\Ws,f)$. In particular, we consider the norm given by an RKHS, a Hilbert space of functions which is associated with a  positive semidefinite (PSD) kernel $\mathcal{K}(x,x')$. Define the matrix $K \in \mathcal{R}^{n \times n}$ as $K_{ij} = K(X_i,X_j)$. Then, we have that
\begin{align*}
\Delta^2(\Ws) &= \sup_{\substack{ \|f\|^2 \leq 1 }} \prns{\nf \sum_{i=1}^n W_i f(X_i,A_i)-\nnf \sum_{i=1}^n \sum_{j=1}^n f(X_i,A_j)}^2 \\
&=\sup_{\substack{ \sum_{i,k=1}^n \alpha_i\alpha_k \K((X_i,X_k),(A_i,A_k)) \leq 1}} \prns{\nf \sum_{i=1}^n W_i f(X_i,A_i) - \nnf \sum_{i=1}^n  \sum_{j=1}^n f(X_i,A_j)}^2 && \text{(a)} \\
&= \sup_{\substack{\alpha^T \K \alpha \leq 1}} \prns{\nf \alpha^\top K \prns{\Ws - \frac{1}{n}e_n}  }^2 \\
&= \| \nf K \prns{\Ws - \frac{1}{n} e_n }\| _2^2  && \text{(b)} \\
&= \nnf \prns{ \prns{\Ws - \frac{1}{n}e_n }^\top K \prns{\Ws - \frac{1}{n}e_n } } \\
&= \nnf \prns{ \Ws^\top K \Ws -\frac{2}{n} e_n^\top K   \Ws  +  \frac{1}{n^2} e_n^\top K e_n },
\end{align*}
\noindent
where $(a)$ is by the representer theorem \citep{scholkopf2001generalized}, and $(b)$ by the dual of the Euclidean norm, and $e_n$ is the vector with $1/n$ in every entry. This results shows that the worst case functional covariance can be expressed as a convex-quadratic function in $\Ws$. To control for extreme weights we propose to add a penalization term and consider the following worst-case penalized functional covariance,
\begin{align}
\notag
\mathfrak{C}(\Ws,\lambda) &= 
\label{worstPFC}
\Delta^2(\Ws)+\frac{\lambda}{n^2}  \| \Ws \|_2^2 .
% \quad CMSE \left(\hat{\tau}^{\op{SATE}}_{W} \right),
\end{align}

When $\lambda$ equals zero, we obtain weights that minimize the covariance. When $\lambda$ is set to be large (depending on the data), we obtain uniform weights. We discuss choices of $\lambda$ in Section \ref{sec:guidelines}. In the next Section, we show how to use quadratic optimization to minimize $\mathfrak{C}(\Ws,\lambda)$.

%%%%%%%%%%%%%%%%%%%%%%%%%%%%%%%%%%%%%%%%%%%%%%%%%%%%%%%%%%%%%%%%%%%%%%%%%%%%%%%%%%%%%%%%%%%%%%%%%%%%%%%%%%%%%%%%%%%%%%%%%%%%%%%%%%%%%%%%%%%%%%%%%%%%%%%%%%%%%%%%%%%%%%%%%%%%%%%%%%%%%%%%%%%%%%%%%%%%%%%%%%%%%%%%%%%%%%%%%%%
%%
%%
%%                  Quadratic optimization to minimize
%%
%%
%%%%%%%%%%%%%%%%%%%%%%%%%%%%%%%%%%%%%%%%%%%%%%%%%%%%%%%%%%%%%%%%%%%%%%%%%%%%%%%%%%%%%%%%%%%%%%%%%%%%%%%%%%%%%%%%%%%%%%%%%%%%%%%%%%%%%%%%%%%%%%%%%%%%%%%%%%%%%%%%%%%%%%%%%%%%%%%%%%%%%%%%%%%%%%%%%%%%%%%%%%%%%%%%%%%%%%%%%%%%%%%%%%%%%%%%%%%%%%%%%%%%%%%%%%%%%%%%%%%%%%%%%%%%%%%%%%%%%%%%%%%%%%%%%%%%%%%%%%%%%%%%%%%%%%%%%%%%%%%%%

\subsection[]{Quadratic optimization to minimize $\mathfrak{C}(\Ws,\lambda)$}
\label{sec:qp_wc}

In the previous Sections we introduced a functional formulation of the covariance between the continuous treatment and the confounders. Since this quantity depends on unknown function we considered the worst-case functional covariance, which is the dual norm of the continuous linear operator, $\delta \prns{\Ws,f}$, showed that it can be expressed as a convex-quadratic function in $\Ws$ and defined the worst case penalized functional covariance, which aim at penalizing extreme weights.  We restric the weights to be positive and sum up to one. Formally, we let $\mathcal W=\fbraces{W_{1:n}\in\R n:W_i\geq0\;\forall i, \sum_{i=1}^n W_i = 1}$. We propose to use weights $\Ws$ obtained by solving the following optimization problem
\begin{equation}
\begin{aligned}
\label{eq:optimization}
\underset{\substack{W_{1:n}\mathcal W}} {\min}\mathfrak{C}(\Ws,\lambda).
\end{aligned}
\end{equation}

As shown in Section \ref{sec:wc_cov_func}, we can express $\mathfrak{C}(\Ws,\lambda)$ as a convex-quadratic function in $\Ws$, and therefore, optimization problem \eqref{eq:optimization} reduces to the following linearly-constrained convex-quadratic optimization problem,
\begin{equation}
\label{qpkom1}
\begin{aligned}
\underset{\substack{W_{1:n}\geq0,\\W_{1:n}^\top e_n=n}}{\min} \frac{1}{n^2} \prns{
\Ws^\top Q\Ws 
-2 \Ws^\top c},
\end{aligned}
\end{equation}

\noindent
where $Q=K+\Sigma_{\lambda}$, $c = e_n^\top K $, and $\Sigma_{\lambda}$ is the diagonal matrix with $\lambda/n^2$ in its $i^{\text{th}}$ diagonal entry.

%%%%%%%%%%%%%%%%%%%%%%%%%%%%%%%%%%%%%%%%%%%%%%%%%%%%%%%%%%%%%%%%%%%%%%%%%%%%%%%%%%%%%%%%%%%%%%%%%%%%%%%%%%%%%%%%%%%%%%%%%%%%%%%%%%%%%%%%%%%%%%%%%%%%%%%%%%%%%%%%%%%%%%%%%%%%%%%%%%%%%%%%%%%%%%%%%%%%%%%%%%%%%%%%%%%%%%%%%%%%%%%%%%%%%%%%%%%%%%%%%%%%%%%%%%%%%%%%%%%%%%%%%%%%%%%%%%%%%%%%%%%%%%%%%%%%%%%%%%%%%%%%%%%%%%%%%%%%%%%%%%%%%%%%%%%%%%%%%%%%%%%%%%%%%%%%%%%%%%%%%%%%%%%%%%%%%%%%%%%%%%%%%%%%%%%%%%%%%%%%%%%%%%%%%%%%%%%%%%%%%%%%%%%%%%%%%%
%%
%%
%%                  Obtaining the dose-response curve
%%
%%
%%%%%%%%%%%%%%%%%%%%%%%%%%%%%%%%%%%%%%%%%%%%%%%%%%%%%%%%%%%%%%%%%%%%%%%%%%%%%%%%%%%%%%%%%%%%%%%%%%%%%%%%%%%%%%%%%%%%%%%%%%%%%%%%%%%%%%%%%%%%%%%%%%%%%%%%%%%%%%%%%%%%%%%%%%%%%%%%%%%%%%%%%%%%%%%%%%%%%%%%%%%%%%%%%%%%%%%%%%%%%%%%%%%%%%%%%%%%%%%%%%%%%%%%%%%%%%%%%%%%%%%%%%%%%%%%%%%%%%%%%%%%%%%%%%%%%%%%%%%%%%%%%%%%%%%%%%%%%%%%%%%%%%%%%%%%%%%%%%%%%%%%%%%%%%%%%%%%%%%%%%%%%%%%%%%%%%%%%%%%%%%%%%%%%%%%%%%%%%%%%%%%%%%%%%%%%%%%%%%%%%%%%%%%%%%%%%

\subsection{Dose-response curve}
\label{sec:dr-curve}

In addition to the data presented in Section \ref{sec:cov_func}, for each unit $i$ in $1,\dots,n$ let $Y_i(a)$ be the potential outcome of treatment $a \in \mathcal A$, and $Y_i=Y_i(A_i)$ the observed outcome. The main object of inference in this paper is the dose-response curve $\theta(a)= \Eb{Y(a)}$. This quantity depends on unknown potential outcomes, and assumptions are needed to identify it in terms of observed data. In this paper we assume consistency, positivity and ignorability \citep{imbens2015causal}. Consistency (together with non-interference) states that the observed outcome equals the potential outcome under the treatment applied to that specific unit, \textit{i.e.}, $Y_i = Y_i(a)$, and that the potential outcomes are well-defined. Positivity states that, the GPS is positive for all values of the confounders in their supports. Ignorability states that, once conditioned on the observed confounders, the potential outcomes are independent to the treatment assignment.   Finally, under these assumption, it can be shown that the dose-response curve $\theta(a)$ is identifiable using observational data.
 
 In the previous Sections, we showed how to obtain weights that minimize the functional covariance defined in \eqref{eq:WFC} by simply solving a linearly-constrained quadratic optimization problem. As described in Section \ref{sec:cov_func}, these weightseliminate any associations between the continuous treatment and the confounders, thus providing optimal covariate balance.
 Once these weights are obtained, to estimate the dose-response curve, we suggest plugging these weights into a weighted parametric or nonparametric estimator. For instance, similar to the approach of \cite{naimi2014constructing,cole2008constructing} for estimating marginal structural models, one can use a weighted nonlinear regression or a weighted local polynomial regression to estimate $\theta(a)$, regressing only the treatment on the outcome. %We provide practical guidelines on its estimation in Section \ref{sec:guidelines}.

%%%%%%%%%%%%%%%%%%%%%%%%%%%%%%%%%%%%%%%%%%%%%%%%%%%%%%%%%%%%%%%%%%%%%%%%%%%%%%%%%%%%%%%%%%%%%%%%%%%%%%%%%%%%%%%%%%%%%%%%%%%%%%%%%%%%%%%%%%%%%%%%%%%%%%%%%%%%%%%%%%%%%%%%%%%%%%%%%%%%%%%%%%%%%%%%%%%%%%%%%%%%%%%%%%%%%%%%%%%%%%%%%%%%%%%%%%%%%%%%%%%%%%%%%%%%%%%%%%%%%%%%%%%%%%%%%%%%%%%%%%%%%%%%%%%%%%%%%%%%%%%%%%%%%%%%%%%%%%%%%%%%%%%%%%%%%%%%%%%%%%%%%%%%%%%%%%%%%%%%%%%%%%%%%%%%%%%%%%%%%%%%%%%%%%%%%%%%%%%%%%%%%%%%%%%%%%%%%%%%%%%%%%%%%%%%%%
%%
%%
%%                  Properties KOOW
%%
%%
%%%%%%%%%%%%%%%%%%%%%%%%%%%%%%%%%%%%%%%%%%%%%%%%%%%%%%%%%%%%%%%%%%%%%%%%%%%%%%%%%%%%%%%%%%%%%%%%%%%%%%%%%%%%%%%%%%%%%%%%%%%%%%%%%%%%%%%%%%%%%%%%%%%%%%%%%%%%%%%%%%%%%%%%%%%%%%%%%%%%%%%%%%%%%%%%%%%%%%%%%%%%%%%%%%%%%%%%%%%%%%%%%%%%%%%%%%%%%%%%%%%%%%%%%%%%%%%%%%%%%%%%%%%%%%%%%%%%%%%%%%%%%%%%%%%%%%%%%%%%%%%%%%%%%%%%%%%%%%%%%%%%%%%%%%%%%%%%%%%%%%%%%%%%%%%%%%%%%%%%%%%%%%%%%%%%%%%%%%%%%%%%%%%%%%%%%%%%%%%%%%%%%%%%%%%%%%%%%%%%%%%%%%%%%%%%%%

\subsection{Properties}
\label{sec:prop}

In this Section, we provide some insights on the properties of dose-response curve estimators weighted by KOOW weights. The overall idea of using weights that minimize the covariance between treatment and confounders has been previously studied in the context of maximal mean discrepancies (MMD) for density-ratio estimation or for testing independence of random variables \citep{sugiyama2012density,gretton2005kernel}. In addition, it is also been shown the connection between MMD and the more general Bregman divergence (see \cite[Section 7.3.2]{sugiyama2012density}). As a consequence, existing results can be used to describe the properties of our proposed methodology. Specifically, by following \cite{sondhi2019balanced,menon2016linking,2019arXiv190101230A}, using kernel functions (as in the local polynomial regression estimator) and under the assumption that exist a set of ``true'' weights that minimize the bias of the weighted estimator, \eg., the stable inverse probability weights, we can bound the bias of the weighted estimator. For instance, the bias of the weighted estimator is bounded by the Bregman distance between the true weights and the weights obtained by solving the optimization problem plus a remainder of a smaller order than the bandwidth parameter when this term goes to infinity \citep[Proposition 2]{sondhi2019balanced}. A similar bound is also provided for the variance \citep[Proposition 3]{sondhi2019balanced}. Furthermore, \cite{sondhi2019balanced} show that under bounded variance, the weighted estimator is also consistent. 

%%%%%%%%%%%%%%%%%%%%%%%%%%%%%%%%%%%%%%%%%%%%%%%%%%%%%%%%%%%%%%%%%%%%%%%%%%%%%%%%%%%%%%%%%%%%%%%%%%%%%%%%%%%%%%%%%%%%%%%%%%%%%%%%%%%%%%%%%%%%%%%%%%%%%%%%%%%%%%%%%%%%%%%%%%%%%%%%%%%%%%%%%%%%%%%%%%%%%%%%%%%%%%%%%%%%%%%%%%%%%%%%%%%%%%%%%%%%%%%%%%%%%%%%%%%%%%%%%%%%%%%%%%%%%%%%%%%%%%%%%%%%%%%%%%%%%%%%%%%%%%%%%%%%%%%%%%%%%%%%%%%%%%%%%%%%%%%%%%%%%%%%%%%%%%%%%%%%%%%%%%%%%%%%%%%%%%%%%%%%%%%%%%%%%%%%%%%%%%%%%%%%%%%%%%%%%%%%%%%%%%%%%%%%%%%%%%
%%
%%
%%                  Practical Guidelines
%%
%%
%%%%%%%%%%%%%%%%%%%%%%%%%%%%%%%%%%%%%%%%%%%%%%%%%%%%%%%%%%%%%%%%%%%%%%%%%%%%%%%%%%%%%%%%%%%%%%%%%%%%%%%%%%%%%%%%%%%%%%%%%%%%%%%%%%%%%%%%%%%%%%%%%%%%%%%%%%%%%%%%%%%%%%%%%%%%%%%%%%%%%%%%%%%%%%%%%%%%%%%%%%%%%%%%%%%%%%%%%%%%%%%%%%%%%%%%%%%%%%%%%%%%%%%%%%%%%%%%%%%%%%%%%%%%%%%%%%%%%%%%%%%%%%%%%%%%%%%%%%%%%%%%%%%%%%%%%%%%%%%%%%%%%%%%%%%%%%%%%%%%%%%%%%%%%%%%%%%%%%%%%%%%%%%%%%%%%%%%%%%%%%%%%%%%%%%%%%%%%%%%%%%%%%%%%%%%%%%%%%%%%%%%%%%%%%%%%%

\section{Practical guidelines}
\label{sec:guidelines}

Solutions to the optimization problem \eqref{eq:optimization} depend on the kernel specification, its hyperparameters and the penalization parameter $\lambda$. In this Section, we provide practical guidelines on their choice. We start by describing the choice of the kernel. Since we want to allow for a great level of flexibility in modeling the relationship between treatment and confounders, we suggest the use of a product of polynomial Mahalanobis kernels:
\begin{equation}\label{polykernel}
\mathcal K((a,x),(a',x'))=\mathcal  K_1(a,a')\mathcal K_2(x,x'),
\end{equation}
\noindent
where
\begin{equation}\label{polykernel}
 \mathcal K_t(z,z')=\gamma_t(1+\theta_t(z-\hat\mu_n)^T\hat\Sigma_n^{-1}(z'-\hat\mu_n))^d,
\end{equation}

\noindent
and where $\gamma_t$ controls the overall scale of the kernel, $\theta_t$ is a parameter that controls the importance of higher orders degrees, $\hat\Sigma_n$ is the sample covariance, $\hat \mu_n$ is the sample mean, and  $d$ is the parameter that controls the degree of the polynomial. \cite{gretton2005kernel} showed that by using universal kernels, such as the Gaussian and Mat\'ern kernels, minimizing \eqref{eq:optimization} leads to statistically independence, whereas by using non-universal kernels, such as polynomial kernels, we aim at uncorrelatedness. Although independence may be preferred  over uncorrelatedness, we find that practically polynomial kernels suffice as shown in our simulations (Section \ref{sec:simu}) and in our illustration (Section \ref{sec:case_study})  . 

\begin{figure}[h!] 
\begin{center}
\includegraphics[scale=.45]{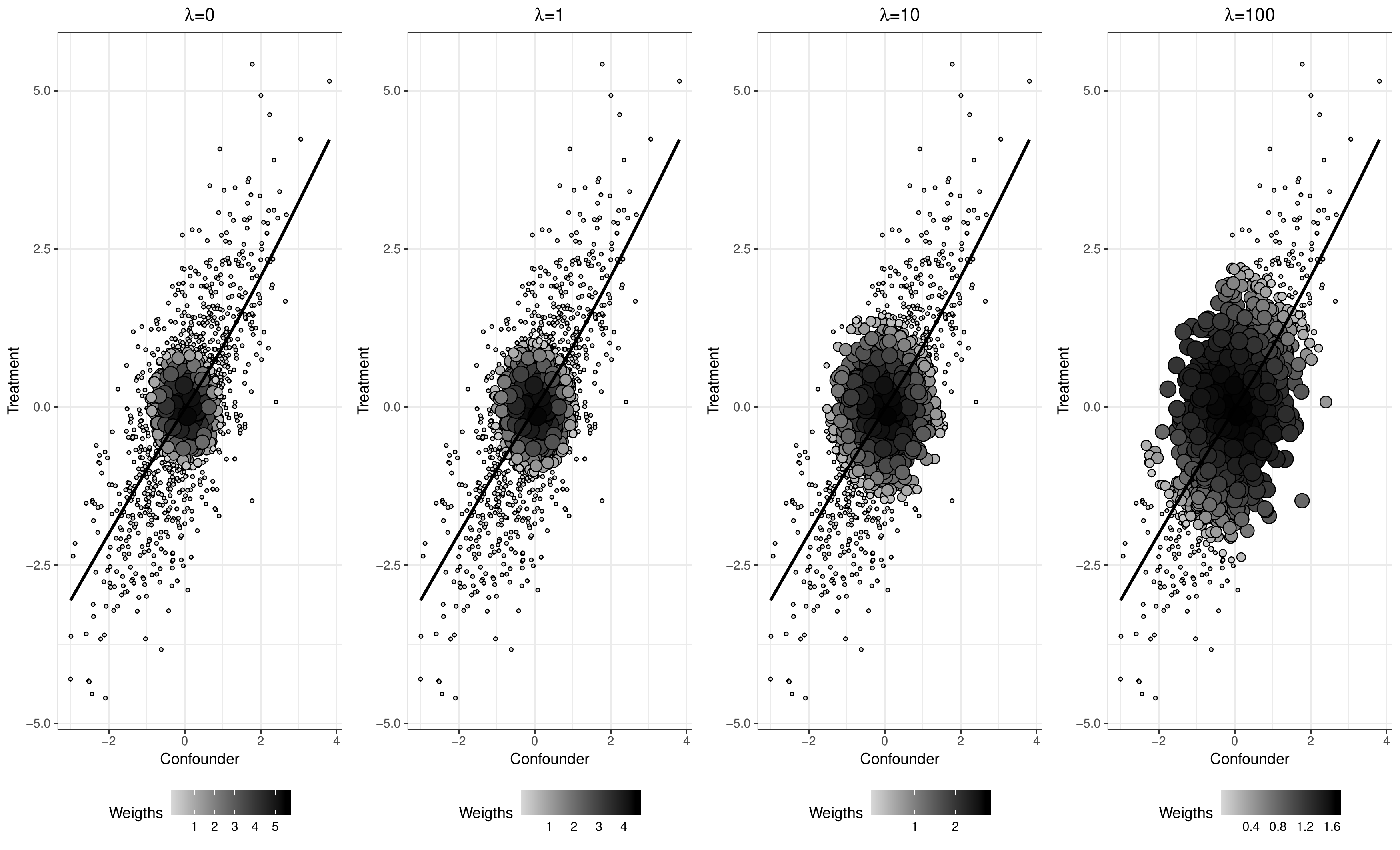}
\end{center}
\caption{\footnotesize The uncorrelatedness-precision trade-off when increasing the value of the penalization parameter $\lambda$ from 0 (first panel from the left) to 100 (last panel from the left). Confounder on the X-axes, Treatment on the Y-Axes.
\label{fig:lambda} }
\end{figure}

We now provide practical guidelines on how to tune the kernel's hyperparameters and the penalization parameter $\lambda$. As shown by \cite{kallus2018more,kallus2018optimal}, we suggest using marginal likelihood, a  model  selection  criteria  for  Gaussian  processes \citep{rasmussen2010gaussian}. To do so, we specify a Gaussian Process (GP) prior, $f$ with covariance identified by the product kernel $\mathcal{K}$, and suppose that we observed the potential outcome $Y_i(a)$ from $f(X_i)$ with Gaussian noise of variance $\sigma^2$. We then maximize the marginal likelihood of seeing the data with respect to the hyperparameters, $\theta_t$, $\gamma_t$, and $\sigma^2$. The penalization parameter $\lambda$ can be interpreted as a uncorrelatedness-precision (bias-variance) trade-off parameter. When set to $0$, the obtained weights targets minimal covariance and therefore minimal bias. When $\lambda$ increases, the obtained weights have lower variance and higher precision can be consequently achieved. An example is provided in Figure \ref{fig:lambda}, in which we plot the scatterplots between a confounder (X-axes) and a treatment (Y-axes) weighted by the set of weights obtained by solving the optimization problem \eqref{eq:optimization}, setting $\lambda$ equal to 0 (first panel from the left of Figure \ref{fig:lambda}), 1 (second panel), 10 (third panel) and 100 (fourth panel). When increasing the penalization parameter $\lambda$, it is clear from this Figure that the distribution of the weights become more uniform (bottom of each scatterplot) thus achieving less uncorrelatedness but increasing precision. In practical setting, an acceptable level of penalization, and therefore of precision, may be for the analyst to determine. We suggest, starting by lower values and, in case, trying to explore different values within reason. We show this in our simulations in Section \ref{sec:simu}.

To estimate standard errors of the dose-response curve, we suggest using bootstrap. Specifically, we suggest to bootstrap the whole process, namely the estimation of the weights, and that of the dose-response curve by a weighted estimator, as shown in our simulation study. 

To tune hyperparameters and solve linearly-constrained convex-quadratic optimization problems several software can be used. We suggest using the \textsf{GaussianProcessRegressor} package from \textsf{scikit-learn} \citep{scikit-learn} for tuning the hyperparameters and \textsf{Gurobi} \citep{optimization2014inc_2} for solving quadratic optimization problems.

%%%%%%%%%%%%%%%%%%%%%%%%%%%%%%%%%%%%%%%%%%%%%%%%%%%%%%%%%%%%%%%%%%%%%%%%%%%%%%%%%%%%%%%%%%%%%%%%%%%%%%%%%%%%%%%%%%%%%%%%%%%%%%%%%%%%%%%%%%%%%%%%%%%%%%%%%%%%%%%%%%%%%%%%%%%%%%%%%%%%%%%%%%%%%%%%%%%%%%%%%%%%%%%%%%%%%%%%%%%%%%%%%%%%%%%%%%%%%%%%%%%%%%%%%%%%%%%%%%%%%%%%%%%%%%%%%%%%%%%%%%%%%%%%%%%%%%%%%%%%%%%%%%%%%%%%%%%%%%%%%%%%%%%%%%%%%%%%%%%%%%%%%%%%%%%%%%%%%%%%%%%%%%%%%%%%%%%%%%%%%%%%%%%%%%%%%%%%%%%%%%%%%%%%%%%%%%%%%%%%%%%%%%%%%%%%%%
%%
%%
%%                  SIMULATIONS
%%
%%
%%%%%%%%%%%%%%%%%%%%%%%%%%%%%%%%%%%%%%%%%%%%%%%%%%%%%%%%%%%%%%%%%%%%%%%%%%%%%%%%%%%%%%%%%%%%%%%%%%%%%%%%%%%%%%%%%%%%%%%%%%%%%%%%%%%%%%%%%%%%%%%%%%%%%%%%%%%%%%%%%%%%%%%%%%%%%%%%%%%%%%%%%%%%%%%%%%%%%%%%%%%%%%%%%%%%%%%%%%%%%%%%%%%%%%%%%%%%%%%%%%%%%%%%%%%%%%%%%%%%%%%%%%%%%%%%%%%%%%%%%%%%%%%%%%%%%%%%%%%%%%%%%%%%%%%%%%%%%%%%%%%%%%%%%%%%%%%%%%%%%%%%%%%%%%%%%%%%%%%%%%%%%%%%%%%%%%%%%%%%%%%%%%%%%%%%%%%%%%%%%%%%%%%%%%%%%%%%%%%%%%%%%%%%%%%%%%

\section{Simulations}
\label{sec:simu}

In this Section, we present the results of a simulation study aimed at evaluating the performance, in terms of integrated absolute bias (IAB) and integrated root mean squared error (IRMSE) across several simulation scenarios. In summary, KOOW performed well with respect to integrated absolute bias and integrated root mean squared error, across the considered scenarios and regardless of the use of a parametric or a nonparametric model for the dose-response curve.

\subsection{Setup}
\label{sec:simu_setup}
%for scenarios (1-2), we generated $$Y_i=0.1+0.5A_i+0.5(X_{1,i}+X_{2,i})+ \beta A_i(X_{1,i}+X_{2,i}),$$ in which $\beta=0$ in (1) and $\beta=0.25$ in (2).

We considered a sample size of $n=1,000$. We generated $$Y_i=0.75A_i+0.05A_i^2+0.01A_i^3+1.5(\sum_{k=1}^5 X_{k,i})+ 1.125 A_i(\sum_{k=1}^5 X_{k,i}),$$ and considered three scenarios for the treatment mechanism: linear, quadratic and cubic. Specifically, we generated $$A_i = \beta_0 + \beta_1 (\sum_{k=1}^5 X_{k,i})^d + \mathcal{N}(0,5)$$ where $\beta_0=(0,-3,-2.5), \beta_1=(1,0.25,0.05)$ and $d=(1,2,3)$, for the linear, quadratic and cubic scenario, respectively, and $X_{k,i} \sim \mathcal{N}(0,5)$, $k=1,\dots,5$.  We generated the true dose-response curves by evaluating the true models at a grid of 1,000 values from -3 to 3. 

We compared KOOW with, outcome modeling (OM) \citep{flores2007estimation}, the non parametric doubly robust estimator of \cite{kennedy2017non} (NPDR), matching on the generalized propensity score (CM) \citep{wu2018matching}, (stable) inverse probability of treatment weighting (IPW) \citep{naimi2014constructing}, parametric (CBPS) and nonparametric covariate balancing propensity score (npCBPS) \citep{fong2018covariate}. We used Super Learner \citep{van2007super} to model the relationships between the outcome the treatment and the confounders (the outcome model) and to model the relationship between the continuous treatment and the confounders (treatment model) for OM, NPDR, and IPW. We considered the following list of algorithms for the Super Learner: linear model, linear model with interactions, linear model with lasso penalization, generalized additive models, multivariate adaptive regression splines, bayesian linear model, and local polynomial regression.  For KOOW, we considered $\lambda=0,1$ and $10$.

We estimated the dose-response curve by using a local polynomial regression estimator with degree 2 and a cubic parametric regression estimator for all methods. Specifically, for KOOW, IPW, CBPS, and npCBPS we plugged the obtained weights into the local polynomial and cubic parametric  regression estimators. For OM, NPDR, and CM, we regressed the treatment on the ``pseudo'' outcomes. We compared KOOW with the other methods with respect to IAB and IRMSE defined as in \cite[Section 4 and Section 5.1, respectively]{kennedy2017non,wu2018matching}. We used \textsf{scikit-learn} (via the \textsf{R} package \textsf{reticulate}) to tune the hyperparametes of the GPs, the \textsf{R} interface of \textsf{Gurobi} to obtain the set of KOOW weights, the \textsf{R} package \textsf{SuperLearner} for computing the outcome and treatment models,  and the \textsf{R} package \textsf{loess} and \textsf{glm} for local polynomial and cubic parametric regression estimation.

\subsection{Results}
\label{sec:simu_res}

In this Section we discuss the results of our simulation study.   In summary, KOOW performed well with respect to IAB and IRMSE across all three considered scenarios and regardeless of the model used to estimate the dose-response curve (local polynomial or cubic regression). Specifically, as shown in Table \ref{tab_simu}, when using a local polynomial regression model, KOOW ($\lambda=0$) obtained the lowest IAB across all scenarios. While OM, NPDR and IPW performed well with respect to IAB and IRMSE under the linear scenario (linear treatment mechanism - first column of Table \ref{tab_simu}), their performance deteriorated when increasing the complexity of the treatment mechanism (second and third columns of Table \ref{tab_simu}). For instance, NPDR had an IAB of 0.49 under the linear scenario, increasing to 3.18 in the quadratic scenario and to 50.58 in the cubic scenario. CM performed relatively well only in the quadratic scenario. CBPS and npCBPS performed well across the three scenarios, suggesting that methods that target balance may have good performance when estimating the effects of continuous treatments. It is worth mentioning that across all scenarios, the IAB of KOOW slightly increased when increasing the penalization parameter $\lambda$, while the IRMSE significantly  decreased. This suggests that KOOW was able to improve precision while introducing negligible bias by penalization.  

We obtained similar results for KOOW, OM, NPDR, IPW and CM when estimating the dose-response curve with a cubic parametric model (Table \ref{tab_simu_p}). CBPS and npCBPS performed worse compared to themselves when estimating the dose-response curve with a local polynomial regression. The IAB of KOOW in the quadratic scenario (second colum of Table \ref{tab_simu_p}) decreased while increasing the penalization parameter $\lambda$ from $0$ to $10$ which is in contrast to the results of all the other scenarios. We argue that this result is related to this specific scenario and  is heuristically explained by the fact that increasing precision by penalization leads to less erroneous estimates and consequently less biased results. Figures \ref{fig:rplot_bias1},\ref{fig:rplot_bias2},\ref{fig:rplot_bias3} in the Supplementary Material show boxplots of the estimated linear (top-left panels), quadratic (top-right panels), and cubic (bottom-left panels) coefficients of the cubic parametric model for the dose-response curve around their true values (horizontal lines) for all the methods and across scenarios (linear, quadratic and cubic). As suggested by the results in Table \ref{tab_simu_p}, KOOW, performed well across scenarios and for all coefficients. Specifically, the interquantile ranges of KOOW's boxplots almost always included the true coefficient. 

\iffalse
Figure \ref{fig:plot_abs_cor} in the Supplementary Material shows that KOOW achieves low MBC, i.e., mean MBC below 0.1 regardless of the value of $\lambda$. CBPS and npCBPS achieved minimal MBC, while IPW and M-GPS performed badly.  Finally, Table \ref{tab_coverage} shows the integrated coverage of the 95\% confidence interval (Ci95CI) of the true dose-response curve, obtained by using the prediction errors from the parametric and nonparametric model for the dose-response curve and that using bootstrap. When using bootstrap prediction errors, the coverages, are close to the nominal value while those obtained from the model are not. We provide additional information about coverage in the plots of Figure \ref{fig:rplot_covefull} in the Supplementary Material. 

\fi

%%%%%%%%%%%%%%%%%%%%%%%%%%%%%%%
%
% Table MISSPECIFIED scenario
%
%%%%%%%%%%%%%%%%%%%%%%%%%%%%%%%

\begin{table}[H]
\centering
 \caption{Integrated Absolute Bias (IAB) and integrated root MSE (IRMSE) across scenarios when estimating the dose-response curve with a local polynomial regression estimator degree 2 \label{tab_simu}}
%\begin{adjustbox}{angle=0,width=\columnwidth,center}
\begin{threeparttable}
\begin{tabular}{lccc}
\hline
                               & \multicolumn{3}{c}{\textbf{Treatment mechanism}}                                 \\ \cline{2-4} 
\multicolumn{1}{l}{\textbf{}} & \textbf{Linear}      & \textbf{Quadratic}   & \multicolumn{1}{l}{\textbf{Cubic}} \\ \cline{2-4} 
                               & \multicolumn{3}{c}{\textbf{IAB (IRMSE)}} \\ \cline{2-4}   
\textbf{Methods}               & \multicolumn{1}{l}{} & \multicolumn{1}{l}{} & \multicolumn{1}{l}{}               \\ \cline{1-1}
\textbf{KOOW $\lambda=0$}      & 0.46 (2.25)          & 0.19 (2.83)          & 0.34 (1.97)                       \\
\textbf{KOOW $\lambda=1$}      & 0.51 (1.93)          & 0.21 (2.25)          & 0.38 (1.75)                        \\
\textbf{KOOW $\lambda=10$}     & 0.67 (1.50)          & 0.23 (1.38)          & 0.59 (1.36)                        \\
\textbf{OM}                    & 1.32 (1.48)          & 4.54 (5.78)          & 17.07 (35.28)                      \\
\textbf{NPDR}                  & 0.49 (0.94)          & 3.18 (4.74)          & 50.58 (157.58)                     \\
\textbf{IPW}                   & 0.50 (2.04)          & 0.37 (1.06)          & 3.87 (4.55)                        \\
\textbf{CM}                    & 9.75 (10.80)         & 1.64 (5.78)          & 27.67 (28.14)                      \\
\textbf{CBPS}                  & 0.67 (4.56)          & 0.44 (0.79)          & 0.56 (2.62)                        \\
\textbf{npCBPS}                & 0.98 (2.61)          & 0.73 (1.00)          & 1.53 (2.43)                        \\ \hline
\end{tabular}
\begin{tablenotes} \footnotesize \item[] Notes: KOOW refers to Kernel Optimal Orthogonality Weighting where $\lambda$ is the penalization parameter; OM refers to Outcome Modeling; NPDR refers to Non Parametric Doubly Robust; IPW to (stable) Inverse Probability Weighting; CM to caliper matching for continuous treatment; CBPS and npCBPS to parametric and non parametric Covariate Balancing Propensity Score. We used Super Learner to model the outcome and treatment models for NPDR, IPW and OM. 
\end{tablenotes}
\end{threeparttable}
%\end{adjustbox}
\end{table}

\begin{table}[H]
\centering
 \caption{Integrated Absolute Bias (IAB) and integrated root MSE (IRMSE) across scenarios when estimating the dose-response curve with a cubic parametric regression \label{tab_simu_p}}
%\begin{adjustbox}{angle=0,width=\columnwidth,center}
\begin{threeparttable}
\begin{tabular}{lccc}
\hline
                               & \multicolumn{3}{c}{\textbf{Treatment mechanism}}                                 \\ \cline{2-4} 
\multicolumn{1}{l}{\textbf{}} & \textbf{Linear}      & \textbf{Quadratic}   & \multicolumn{1}{l}{\textbf{Cubic}} \\ \cline{2-4} 
                               & \multicolumn{3}{c}{\textbf{IAB (IRMSE)}} \\ \cline{2-4}   
\textbf{Methods}               & \multicolumn{1}{l}{} & \multicolumn{1}{l}{} & \multicolumn{1}{l}{}               \\ \cline{1-1}
\textbf{KOOW $\lambda=0$}      & 0.46 (2.44)          & 0.28 (4.88)          & 0.13 (2.23)                       \\
\textbf{KOOW $\lambda=1$}      & 0.51 (2.10)          & 0.10 (3.46)          & 0.15 (1.95)                        \\
\textbf{KOOW $\lambda=10$}     & 0.71 (1.61)          & 0.07 (1.95)          & 0.27 (1.48)                        \\
\textbf{OM}                    & 2.93 (3.00)          & 4.53 (5.78)          & 20.26 (36.01)                      \\
\textbf{NPDR}                  & 0.14 (1.00)          & 1.01 (5.92)          & 7.36 (72.98)                     \\
\textbf{IPW}                   & 1.77 (3.36)          & 0.89 (3.94)          & 3.85 (4.90)                        \\
\textbf{CM}                    & 1.55 (8.41)         & 4.99 (34.48)          & 42.80 (44.92)                      \\
\textbf{CBPS}                  & 2.51 (5.46)          & 3.24 (7.04)          & 3.15 (5.81)                        \\
\textbf{npCBPS}                & 6.98 (7.67)          & 1.51 (7.16)          & 12.45 (12.66)                        \\ \hline
\end{tabular}
\begin{tablenotes} \footnotesize \item[] Notes: KOOW refers to Kernel Optimal Orthogonality Weighting where $\lambda$ is the penalization parameter; OM refers to Outcome Modeling; NPDR refers to Non Parametric Doubly Robust; IPW to (stable) Inverse Probability Weighting; CM to caliper matching for continuous treatment; CBPS and npCBPS to parametric and non parametric Covariate Balancing Propensity Score. We used Super Learner to model the outcome and treatment models for NPDR, IPW and OM. 
\end{tablenotes}
\end{threeparttable}
%\end{adjustbox}
\end{table}

\section{Case-study}
\label{sec:case_study}

In this Section, we apply KOOW to the evaluation of the effect of red meat consumption on blood pressure among women of the Women's Health Initiative (WHI) observational study. 

\subsection{The effect of red meat consumption on hypertension among women of the WHI observational study}

Despite increasing efforts to early detect and treat high blood pressure in recent years, hypertension still remains one of the major risk factor for stroke, cardiovascular diseases and mortality worldwide \citep{wang2004prevalence,lawes2006blood,collaboration2002age}. In the eighties, the ``iron-heart'' hypothesis \citep{sullivan1981iron} was proposed to explain possible differences in rates of coronary heart disease between pre-menopausal women and men. This hypothesis, was based on the idea that pre-menopausal women loose iron through menstruation. Since these results, several studies have evaluated the relationship between iron and coronary heart disease suggesting that greater haem iron (the iron originated from animal sources) intake increases the risk of coronary heart disease \citep[for a recent meta-analysis]{wang2016red}. Red meat is a major source of haem iron and several studies and guidelines suggest consuming a modest amount of red meat.  Most of these studies categorize red meat consumption in two or few categories and evaluate their effects on a binary outcome, such as presence or not of hypertension (see \cite{kappeler2013meat} for example). However, as described in Section \ref{sec:intro}, a more natural object of interest is the dose-response curve, which directly evaluates the amount of red meat that a person eats every day on a continuous outcome. Therefore, in this Section we apply KOOW to estimate the dose-response curve of red meat consumption and systolic blood pressure.

\subsection{Study population}

We used a subset of 2,000 randomly selected units from the the Women’s Health Initiative (WHI) observational study \citep{study1998design}. WHI is one of the largest cohort of postmenopausal women aged 50 to 79 years that provides valuable information about demographic and clinical information such as data on blood pressure and hormone therapy treatment. 
Red meat consumption was recorded as estimated medium servings per day. We restricted our analyses to those women between the 5th and the 95th quantile of the distribution of red meat consumption. Thus, the minimum medium red meat serving per day in our subset was equal to 0.04, the mean 0.53, and the maximum 1.51.  We identified as potential confounders the following variables: systolic blood pressure at baseline, age, total dietary energy intake (measure in kcal), body mass index (BMI) (calculated as weight in kilograms divided by the square of height in meters), smoking status (defined as if a woman ever smoked at least 100 cigarettes), and hormone therapy (HT) ever used (defined as if a woman ever used HT).

\subsection{Model setup}

We obtained the set of KOOW weights by solving optimization problem \ref{eq:optimization}. We set $\lambda$ equal to 10. We chose a product of two polynomial kernel degree 1, and we tuned their hyperparameters by using GPML as described in Section \ref{sec:simu}. We estimated the dose-response curve by using a local polynomial regression estimator degree 1 weighted by the set of KOOW weights. We bootstrapped the whole process of obtaining the KOOW weights and estimating the dose-response curve, 1000 times. For each bootstrap sample, we computed the predicted dose-response curve and finally estimated its bootstrap confidence interval.  

\subsection{Results}

\begin{figure}[H] 
\begin{center}
%\begin{adjustbox}{angle=0,width=\columnwidth,center}
\includegraphics[scale=1]{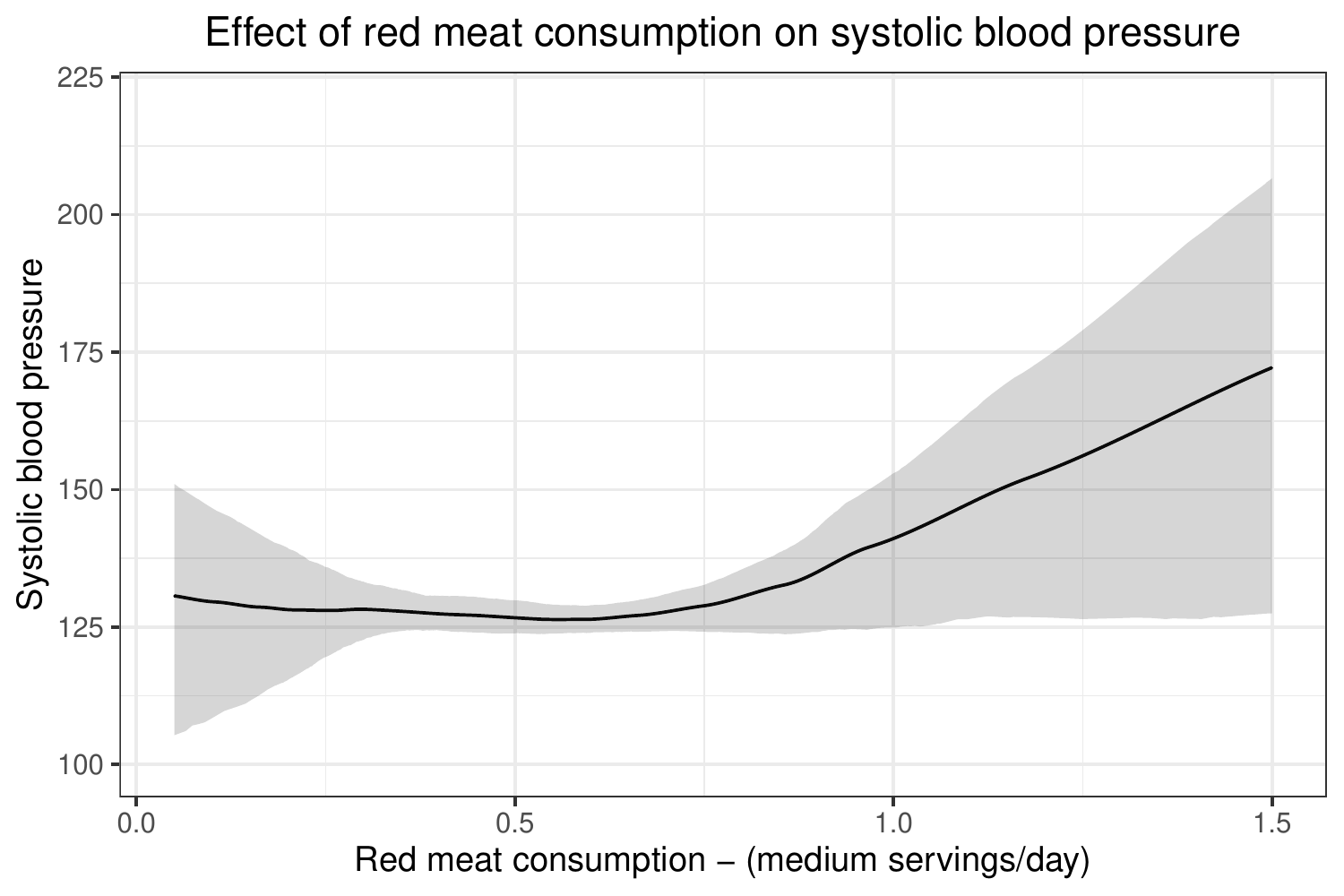}
%\end{adjustbox}
\end{center}
\caption{\footnotesize Estimated dose-response curve between red meat consumption (x-axis) and Systolic blood pressure (y-axis). Confidence intervals are obtained by bootstrap (1,000 replications).
\label{fig:rplot_cs0} }
\end{figure}

\begin{figure}[H] 
\begin{center}
%\begin{adjustbox}{angle=0,width=\columnwidth,center}
\includegraphics[scale=0.7]{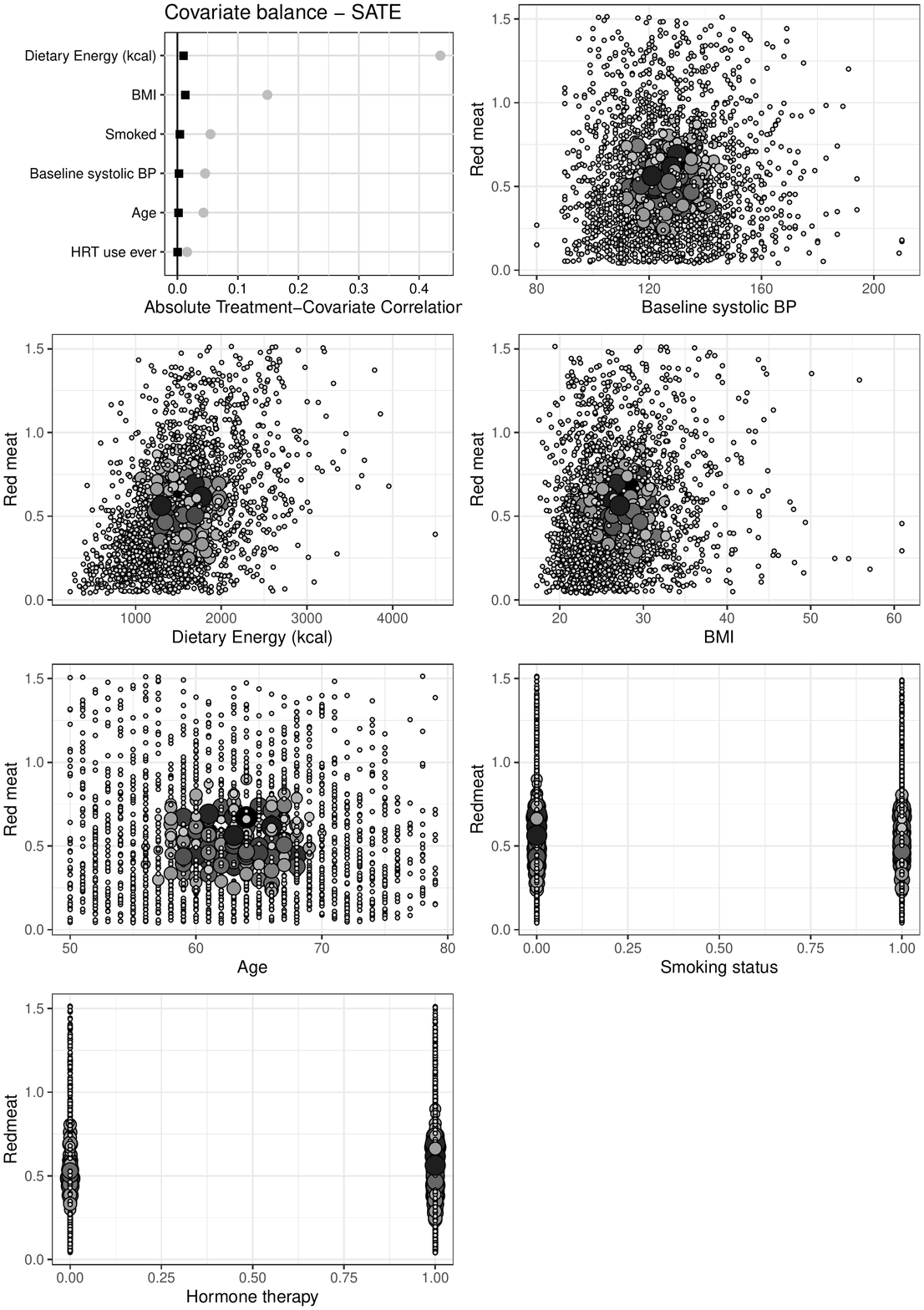}
%\end{adjustbox}
\end{center}
\caption{\footnotesize Top left panel: absolute weighted correlation between the continuous treatment and the confounders. Other panels show the relationship between the confounders (baseline systolic blood pressure, dietary energy, BMI, age, smoking status and HT) and red meat consumption, reweighted by the set of KOOW weights (size and color represent the weights, larger and darker units represent higher weights).
\label{fig:rplot_cs1} }
\end{figure}

In this Section we present the results of our analysis. Similar to the results of previous studies we show a possible association between red meat consumption and hypertension. Figure \ref{fig:rplot_cs0}, shows the dose response curve between red meat consumption (x-axis) and the systolic blood pressure (y-axis). Mean systolic blood pressure was around 125 for those women with low medium red meat consumption, while increasing, up to almost 175, along with the consumption of red meat. The top left panel of Figure \ref{fig:rplot_cs1} shows adjusted, \ie., weighted by the KOOW weights, covariate balance (black dots) and unadjusted (grey dots) of the six confounders considered in our analysis. Covariate balance was computed as the absolute weighted correlation between the continuous treatment and the confounders. KOOW shows to minimize covariate balance to values close to 0. Figure \ref{fig:rplot_cs1} also shows the scatterplots between each of the six considered confounders (x-axes of the top right panel and the middle and bottom panels of Figure \ref{fig:rplot_cs1}), and red meat consumption (y-axes). As shown in Figure \ref{fig:plot_expw} in Section \ref{sec:cov_func}, KOOW provide weights that make confounders and red meat consumption orthogonal.  For completeness we also report the weighted units for the binary confounders smoked and hormone therapy in the bottom panels of Figure   \ref{fig:rplot_cs1}. Finally, based on the results of our analysis and similar to previous epidemiological and medical literature, we conclude that there may be a possible positive relationship between red meat consumption and systolic blood pressure.

%%%%%%%%%%%%%%%%%%%%%%%%%%%%%%%
%
% Figure Bias (plot_bias.R)
%
%%%%%%%%%%%%%%%%%%%%%%%%%%%%%%%

\section{Conclusion}
\label{sec:conc}

Continuous treatments arises very often in practice and proper estimation of dose-response curves is crucial for medical decision making. In this paper we proposed a novel method based on quadratic optimization that finds weights that minimize the covariance between the continuous treatment and the confounders, thus eliminating any possible relationships between the two and consequently optimizing covariate balance. 

%Unlike traditional methods based on the GPS, 
%By using mathematical optimization, KOOW provides a flexible way to find weights that control for confounding to estimate effects of continuous treatments while simultaneously controlling for extreme weights. 
By using kernels, the proposed method automatically learns the structure of the data without relying on parametric assumptions about the outcome model, which, in most practical settings, are hardly ever met. In addition, as shown in Section \ref{sec:guidelines}, the penalization parameter $\lambda$ can be easily tuned and interpreted as the uncorrelatedness-precision (bias-variance) trade-off parameter. The proposed method provided low integrated absolute bias and low integrated RMSE across several different scenario, as showed in our simulation scenarios, 

In this paper, we tuned the kernels' hyperparametes by using Gaussian Process Marginal Likelihood. Other methods may be also used. For instance, one can use a kernelized regression and tune the kernels hyperparameters by usual cross validation techniques. 

In this paper, once the set of weights was obtained, we plugged it into a weighted estimator for the dose-response curve. Alternative weighted estimation techniques may also be considered. For instance, one may first estimate the KOOW weights and the plug them into a parametric or nonparametric augmented estimator, similar to that of \cite{kennedy2017non}. Another option would be to use a KOOW-weighted regression estimator that includes both treatment and confounders as in \cite[Section 3.2]{kang2007demystifying}.

%Possible extension of KOOW in the setting of longitudinal data, in which methods must control for time-dependent confounding, confounders that are affected by previous treatments and affect future ones, is also of interest. 

\bibliographystyle{chicago}

\bibliography{bib}

\newpage
\bigskip
\begin{center}
{\large\bf SUPPLEMENTARY MATERIAL}
\end{center}

\begin{figure}[b] 
\begin{center}
\begin{adjustbox}{angle=0,width=\columnwidth,center}
\includegraphics[scale=1]{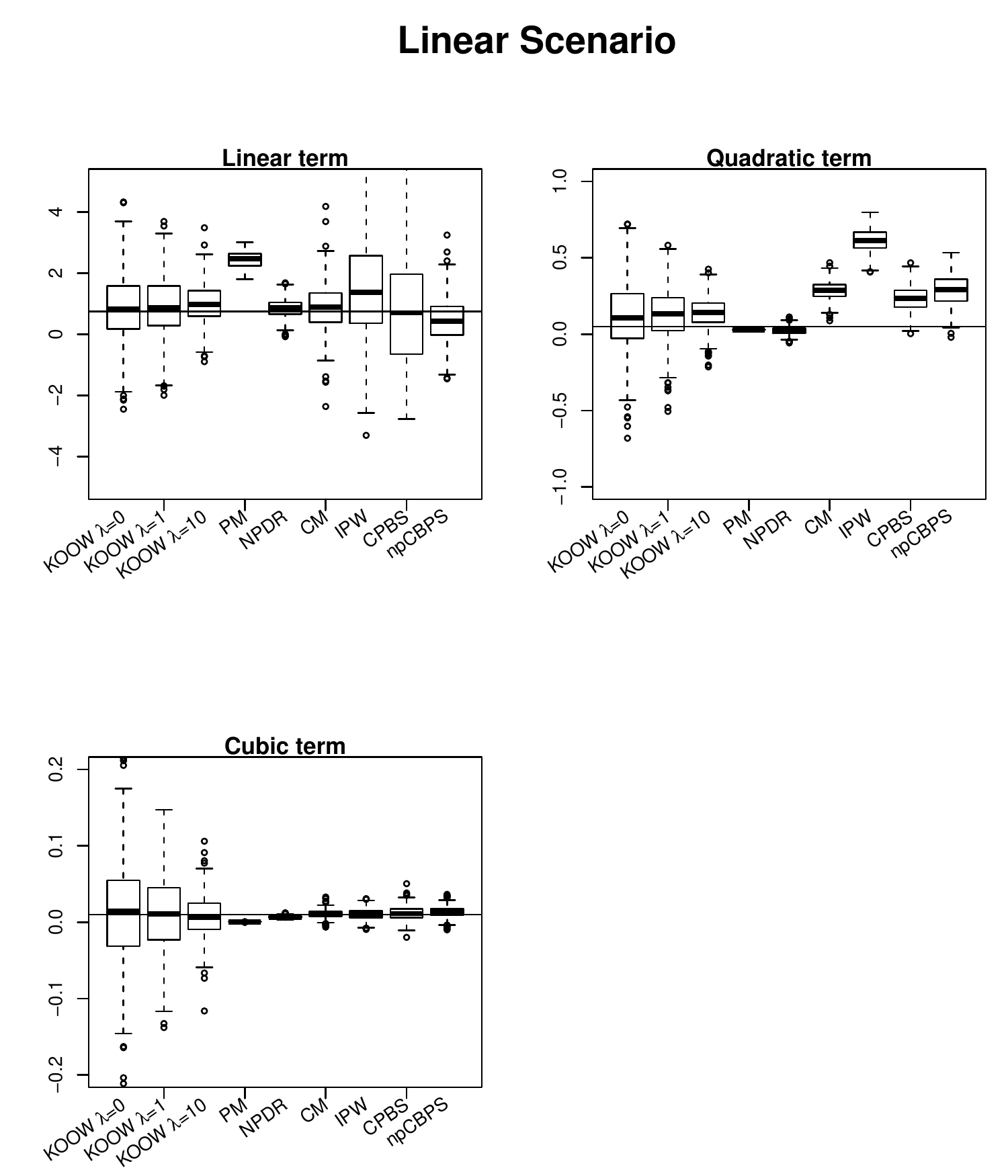}
\end{adjustbox}
\end{center}
\caption{\footnotesize \textit{Linear Scenario:} Boxplots  of  the  estimated  linear  (top-left panel),  quadratic  (top-right  panel),  and  cubic (bottom-left panel) coefficients of the cubic parametric model for the dose-response curve around  their  true  values  (horizontal  lines)  for  all  the  methods  under the linear scenario.
\label{fig:rplot_bias1} }
\end{figure}

\begin{figure}[b] 
\begin{center}
\begin{adjustbox}{angle=0,width=\columnwidth,center}
\includegraphics[scale=1]{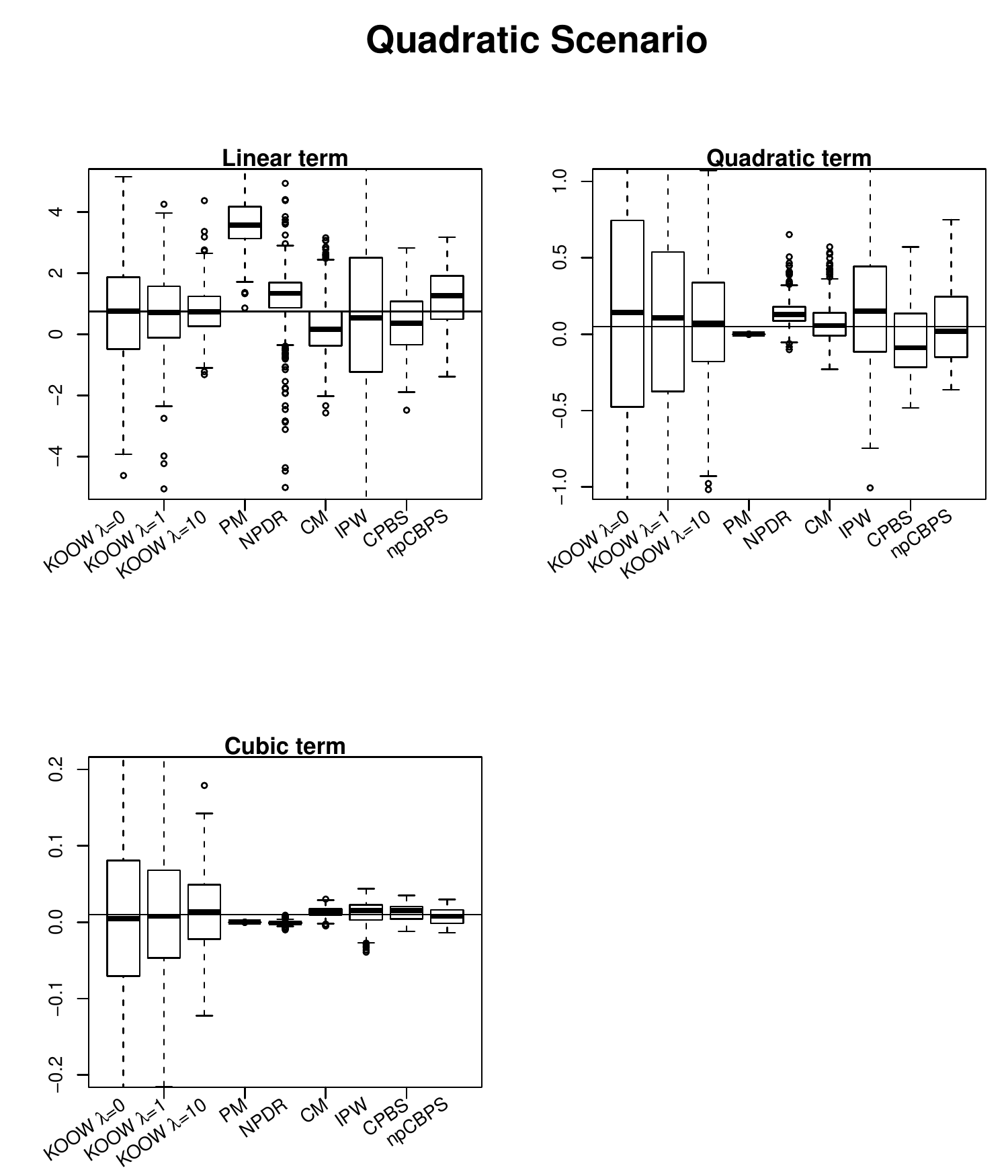}
\end{adjustbox}
\end{center}
\caption{\footnotesize \textit{Quadratic Scenario:} Boxplots  of  the  estimated  linear  (top-left panel),  quadratic  (top-right  panel),  and  cubic (bottom-left panel) coefficients of the cubic parametric model for the dose-response curve around  their  true  values  (horizontal  lines)  for  all  the  methods  under the quadratic scenario.
\label{fig:rplot_bias2} }
\end{figure}

\begin{figure}[b] 
\begin{center}
\begin{adjustbox}{angle=0,width=\columnwidth,center}
\includegraphics[scale=1]{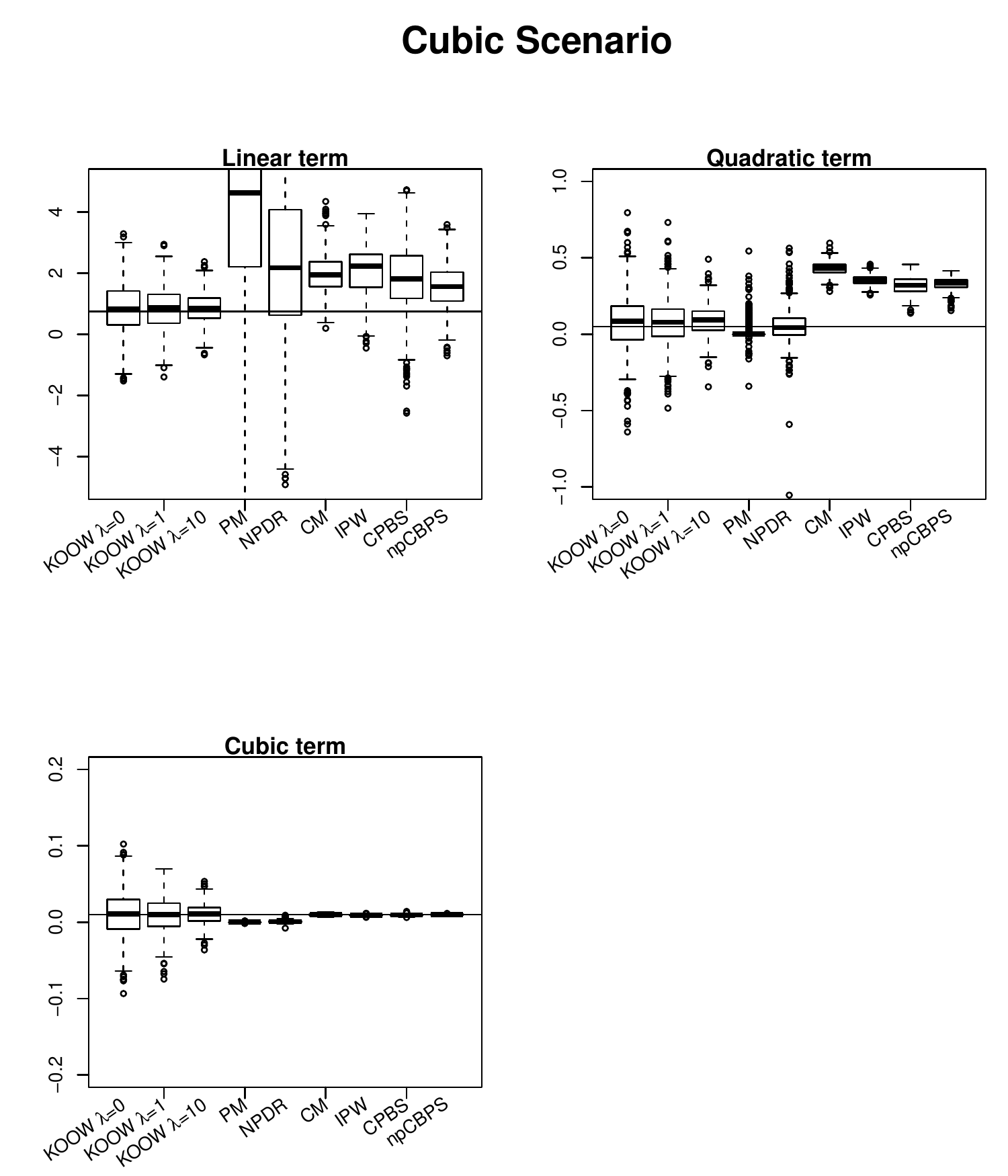}
\end{adjustbox}
\end{center}
\caption{\footnotesize \textit{Cubic Scenario:} Boxplots  of  the  estimated  linear  (top-left panel),  quadratic  (top-right  panel),  and  cubic (bottom-left panel) coefficients of the cubic parametric model for the dose-response curve around  their  true  values  (horizontal  lines)  for  all  the  methods  under the cubic scenario.
\label{fig:rplot_bias3} }
\end{figure}

\begin{description}

\item[Simulations:] Additional Figures\ref{fig:rplot_bias1},\ref{fig:rplot_bias2},\ref{fig:rplot_bias3} showing the boxplots of the estimated linear (top-left panels),quadratic (top-right panels), and cubic (bottom-left panels) coefficients of the cubic para-metric model for the dose-response curve around their true values (horizontal lines) for all the methods and across scenario.

\end{description}

\end{document}